%% file: main.tex
\documentclass[pdflatex,sn-nature]{sn-jnl}
\usepackage{graphicx}
\usepackage{amsmath,amssymb}
\usepackage{booktabs}
\usepackage{xcolor}
\usepackage{placeins}
\usepackage{siunitx}
\usepackage{url}
\usepackage{bookmark}

\setcitestyle{super,open={},close={}}

\input{layout_tune.tex}
\input{si-xr.tex}
\input{gnn_numbers.tex}
\input{gen_dft_numbers.tex}

\AtBeginDocument{\hypersetup{colorlinks=true,
  linkcolor={blue!55!black},citecolor={blue!55!black},
  urlcolor={blue!55!black},breaklinks=true,bookmarksnumbered=true,pdfstartview={FitH}}}
\newcommand{\alttext}[1]{}
\newcommand{\tablefont}{\scriptsize\linespread{1.0}\selectfont\setlength{\tabcolsep}{3pt}}
\begin{document}
\input{frontmatter.tex}

\author[1,2]{\fnm{Zhilong} \sur{Song}}
\author[1]{\fnm{Zijun} \sur{Pang}}
\author*[1,2,3]{\fnm{Lixue} \sur{Cheng}}\email{lixuecheng@ust.hk}
\affil[1]{Department of Chemistry, Hong Kong University of Science and Technology, Kowloon, Hong Kong 999077, China}
\affil[2]{IAS Center for AI for Scientific Discoveries, Hong Kong University of Science and Technology, Kowloon, Hong Kong 999077, China}
\affil[3]{Department of Chemical and Biological Engineering, Hong Kong University of Science and Technology, Kowloon, Hong Kong 999077, China}
\maketitle
\clearpage
\section{Introduction}\label{sec:introduction}
\input{intro.tex}
\section{Results}
\input{results_record.tex}

\input{results_labels.tex}

\input{results_completion.tex}

\input{results_novelty.tex}
\input{discussion.tex}
\input{methods.tex}
\backmatter
\bookmarksetup{startatroot}
\input{availability.tex}

\begingroup
\linespread{1.60}\selectfont
\bibliography{refs}
\endgroup
\input{backmatter.tex}
\end{document}

%% file: layout_tune.tex
\AtBeginDocument{\setlength{\bibsep}{1pt plus 0.3pt}}

\makeatletter
\long\def\@makecaption#1#2{%
  \vskip\abovecaptionskip
  \begingroup
    \small \linespread{1.0}\selectfont
    \leftskip\z@ \rightskip\z@ \parfillskip\z@ plus 1fil\relax
    \sbox\@tempboxa{\textbf{#1}\quad #2}%
    \ifdim \wd\@tempboxa >\hsize
      \textbf{#1}\quad #2\par
    \else
      \hb@xt@\hsize{\textbf{#1}\quad #2\hfil}%
    \fi
  \endgroup
  \vskip\belowcaptionskip}
\makeatother

\makeatletter
\renewcommand{\Titlefont}{\reset@font\linespread{1.0}\fontsize{17bp}{22.5bp}\selectfont\centering}
\makeatother

%% file: gnn_numbers.tex
\newcommand{\NbenchWins}{nine}

\newcommand{\FormationMeV}{23.7}

\newcommand{\MinJarvisEntries}{40}
\newcommand{\WithinAgreeWindow}{89.0}
\newcommand{\WithinWorstPair}{99.0}

%% file: gen_dft_numbers.tex
\providecommand{\dftex}[1]{#1}
\newcommand{\dftAN}{\dftex{98}}

\newcommand{\dftAOsPhrase}{\dftex{all}}

\newcommand{\dftAbest}{\dftex{401}}
\newcommand{\dftAchem}{\dftex{5}}

\newcommand{\dftAhits}{\dftex{98}}

\newcommand{\dftArecorded}{\dftex{80}}

\newcommand{\dftBN}{\dftex{94}}

\newcommand{\dftBbest}{\dftex{440}}

\newcommand{\dftBexpN}{\dftex{85}}

\newcommand{\dftBexpSN}{\dftex{77}}

\newcommand{\dftBexpSchem}{\dftex{39}}
\newcommand{\dftBexpShits}{\dftex{56}}

\newcommand{\dftBexpchem}{\dftex{25}}

\newcommand{\dftBexphits}{\dftex{63}}

\newcommand{\dftBhits}{\dftex{55}}

\newcommand{\dftReNtwo}{\dftex{380}}

\newcommand{\dftSallN}{\dftex{78}}

\newcommand{\dftSallmae}{\dftex{14.6}}

\newcommand{\dftShiN}{\dftex{93}}
\newcommand{\dftShihits}{\dftex{54}}
\newcommand{\dftShirate}{\dftex{58}}
\newcommand{\dftTfourN}{\dftex{379}}
\newcommand{\dftTfourbiref}{\dftex{263}}
\newcommand{\dftTfourpass}{\dftex{211}}
\newcommand{\dftTfourpreN}{\dftex{32}}
\newcommand{\dftTfourprebiref}{\dftex{23}}
\newcommand{\dftTfourprepass}{\dftex{16}}
\newcommand{\dftTonepre}{\dftex{27}}
\newcommand{\dftTonepredone}{\dftex{30}}
\newcommand{\dftTonesynth}{\dftex{38}}
\newcommand{\dftTonesynthB}{\dftex{46}}
\newcommand{\dftTonesynthN}{\dftex{57}}
\newcommand{\dftTonetop}{\dftex{21}}
\newcommand{\dftTonetopG}{\dftex{17}}
\newcommand{\dftTonetopdone}{\dftex{27}}
\newcommand{\msTonepre}{29}

%% file: frontmatter.tex
\title{Probing the limits of materials design\\
via a comprehensive crystal database of\\
14.6 million structures with 140 properties}

\abstract{Forward screening and inverse design drive materials discovery, yet their limits remain unquantified because records of known crystal structures and properties are fragmented across databases.
Consequently, forward screening leaves known crystals underexplored for new functions, whereas generative and autonomous discovery has repeatedly reported known phases or close variants as new materials.
Here we match structures across 55 databases to assemble MateriaViva, a unified database of 14.6 million distinct crystals and over one billion property values.
To provide the missing properties of every structure, a task we term property completion, we develop an equivariant Crystal Graph Transformer Network with state-of-the-art accuracy on \NbenchWins{} benchmark tasks and apply 140 property-prediction models across the database.
These predictions reveal the underestimated potential of forward screening, identifying tens of thousands of high-performance candidates among previously synthesized or computationally studied materials that lack recorded target-property values.
Conversely, comparison with MateriaViva shows that the novelty of inverse design has been overestimated, as 55--100\% of inverse-designed structures are not new.
Under equal density functional theory verification budgets for bulk-modulus design, inverse design yields a higher fraction of qualifying candidates, whereas forward screening reaches the higher bulk modulus.
MateriaViva thus provides a comprehensive reference for probing the limits of both paradigms, revealing unrecognized performance in known crystals and directing inverse design beyond reported structures.}

\keywords{crystalline materials design, high-throughput screening, inverse design,
crystal structure deduplication, provenance, machine learning for materials}

%% file: intro.tex
The search for crystalline materials with targeted properties relies on two
complementary paradigms, forward screening and inverse design. Forward screening
evaluates known crystals with high-throughput first-principles calculations or
accurate property-prediction models to select candidates for a target application
\cite{ceder1998identification,curtarolo2013high,xie2018crystal,mounet2018two,zhang2019catalogue,shamshirgar2026data},
whereas inverse design generates crystal structures with specified target
properties \cite{zunger2018inverse,merchant2023scaling,szymanski2023autonomous,zeni2025generative}.
Because screening draws its candidates from an existing collection of structures
while inverse design can propose structures beyond that collection
\cite{szymanski2025baselines}, a fundamental question arises: what limits each
paradigm, and does generating new structures necessarily deliver better
materials than screening those already known?

The potential of forward screening may be underestimated when known crystal
structures are treated as materials whose useful properties have already been
explored. Screening of known compounds has repeatedly revealed opportunities
that their original characterization did not identify. A screen of
108{,}423 experimentally known compounds, for example, uncovered 1{,}825
that could be readily or potentially exfoliated into two-dimensional materials
\cite{mounet2018two}, while a survey of 39{,}519 compounds in crystal databases
catalogued 8{,}056 with nontrivial electronic topology \cite{zhang2019catalogue}.
More recently, machine-learning-assisted database mining recognized 33 previously
synthesized compounds as unrecognized candidates for MXenes, a family of
two-dimensional transition-metal carbides and nitrides
\cite{shamshirgar2026data}. These studies show that evaluating an unreported
property or recognizing an overlooked function can turn a known crystal into
a candidate for a new application. The potential of screening therefore
depends on how fully the properties of known crystals have been explored.
This raises the complementary question of whether inverse design genuinely reaches beyond those known crystals.

For inverse design, incomplete coverage of known crystals affects novelty assessment. Novelty is generally assessed by comparing generated structures
with crystals recorded in a few large databases
\cite{song2026nsr,lu2026breaking,ramlaoui2025lemat,betala2025lematgenbench}. A known
structure can therefore be counted as new when the database that contains it is
excluded from the comparison. A survey of GNoME's 384{,}870 stable predictions \cite{merchant2023scaling} found many of
them to be elemental substitutions on known structure types
\cite{cheetham2024artificial}. When the
autonomous laboratory A-Lab reported the synthesis of dozens of new compounds \cite{szymanski2023autonomous}, a reanalysis concluded that
none was new. In many cases, the proposed ordered structures assigned elements to distinct
crystallographic sites, whereas the experimental diffraction patterns were
consistent with known disordered phases in which those elements share sites
\cite{leeman2024challenges}. MatterGen's flagship compound, synthesized to
demonstrate property-targeted generation, was later identified as a disordered
phase reported five decades earlier and present in the model's own training set
\cite{juelsholt2026continued}. An ion-exchange baseline built from known
compounds also outperformed four generative techniques at producing stable
crystals \cite{szymanski2025baselines}. In the GNoME, A-Lab and MatterGen cases,
structures reported as new were already recorded or could be derived from known
structures, but comparisons with the selected databases failed to identify them.
Evaluating screening and generation therefore requires establishing which
crystal structures and properties are already known. Linking the records of
each crystal across databases provides the structural and property information
needed for this evaluation.

Building such a unified reference is challenging because millions of
experimental and computational crystal records, with their measured and
calculated property values, are fragmented across separate repositories
\cite{jain2013commentary,saal2013materials,curtarolo2012aflow,%
schmidt2022dataset,choudhary2020jarvis,draxl2019nomad,talirz2020materials,%
zagorac2019recent,grazulis2009crystallography}.
Although these collections underpin data-driven crystal discovery and generation
\cite{merchant2023scaling,xie2022crystal,antunes2024crystal,zeni2025generative}
and broadly applicable interatomic potentials
\cite{chen2022universal,deng2023chgnet,batatia2023foundation,yang2024mattersim,wood2025uma},
even large molecular-dynamics (MD) datasets such as OMat24 contain distinct crystal
structures only on the order of $10^5$ \cite{barrosoluque2026omat24,gharakhanyan2026omc25},
leaving most known crystals scattered across specialized databases. Unifying these
sources requires a common crystal identifier, because accession numbers differ
across repositories and chemical formulae cannot distinguish polymorphs. While
OPTIMADE standardizes cross-database queries \cite{andersen2021optimade} and
LeMaterial removes duplicate structures across several large computational
databases \cite{ramlaoui2025lemat}, unifying the full range of experimental and
computational databases also requires reconciling heterogeneous property
definitions, since two fields labelled band gap, for example, may report different
physical quantities or levels of theory. Moreover, even after structural and
property integration, most crystal--property pairs remain empty because the
original studies measured or calculated only selected properties. Predicting these
missing values, which we term property completion, makes known crystals searchable
for applications absent from their original records. It requires models that
remain accurate across hundreds of tasks, often with only a few thousand labels,
and on chemical systems held out from training. Crystal-level integration and
property completion thus provide the two foundations needed to evaluate the
novelty of generated crystals and to search known crystals for new applications.

Here we assemble MateriaViva, a unified collection
of 55 crystal databases spanning general-purpose repositories, experimental
structure databases and specialized materials collections, to probe and extend the
limits of forward screening and inverse design. Matching structures across these
sources yields 14{,}590{,}485 distinct crystals and 131.8 million values for 166
properties, each retaining its source and calculation or measurement method, yet
only 4.5\% of crystal--property pairs have recorded values. To fill these gaps, we
developed CGTNet (Crystal Graph Transformer Network), a rotation-equivariant graph
transformer with state-of-the-art accuracy on \NbenchWins{} benchmark tasks, and
applied 140 property-prediction models across the database (released as an open
model library on Hugging Face), adding 874.1 million predicted property values. Screening the completed database for four applications,
namely ultra-incompressible solids, Li-ion conductors, high-voltage cathodes and
ultraviolet (UV) birefringent crystals, identifies 62{,}473 candidates without a
recorded target-property value, including 580 with synthesis evidence. Among them is
ReN$_2$, a phase synthesized at high pressure with a measured bulk modulus of
428~GPa \cite{bykov2019rhenium}, for which none of the 55 databases records a bulk
modulus. Conversely, comparison with MateriaViva shows that 55--100\% of
inverse-designed structures are not new, matching known crystals in composition or
in both stoichiometry pattern and space group. Under equal density functional
theory (DFT) verification budgets for bulk-modulus design, inverse design yields a
higher fraction of mechanically stable candidates with bulk moduli of at least
250~GPa, whereas forward screening reaches the higher bulk modulus and its
experimental-source subsets span more chemical systems. By making unreported properties searchable, establishing which structures are
already known and self-expanding through an autonomous database-ingestion agent,
MateriaViva reveals unrecognized functionality within known crystals and directs
inverse design towards structures beyond those already reported.

%% file: results_record.tex
\subsection{Integration of known crystals}
\label{sec:index}

\begin{figure*}
  {\centering\includegraphics{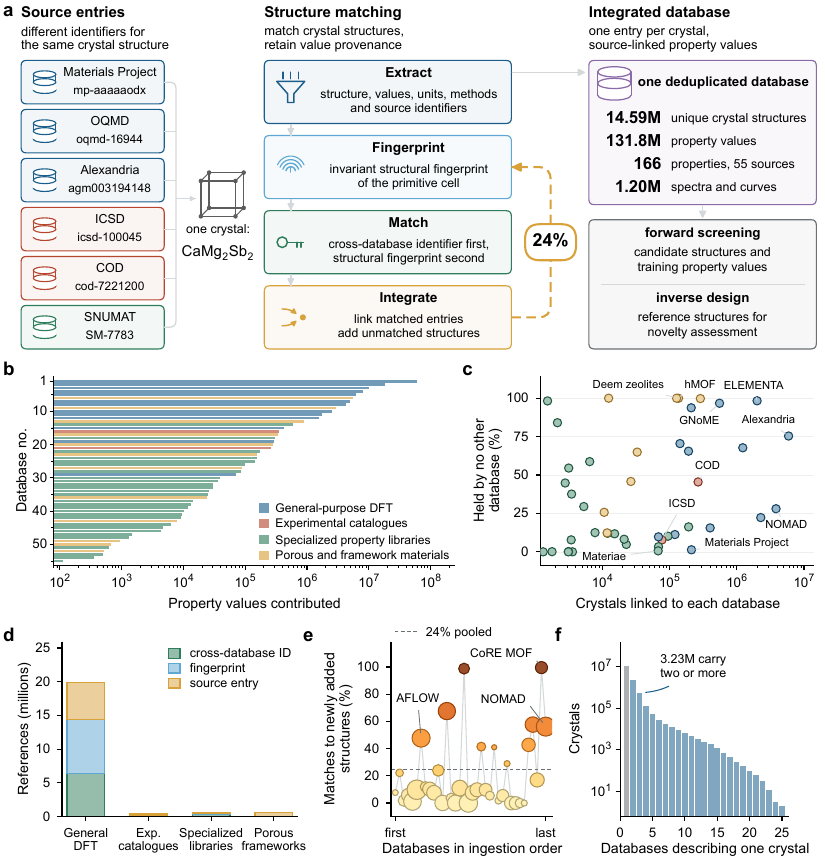}\par}
  \caption{\textbf{Integration of crystal structures and property data across databases.}
  (a) Matching workflow illustrated with the published identifiers of
  CaMg$_2$Sb$_2$. Unmatched structures are added to the reference set (dashed
  bracket). The percentage gives the fraction of all resolved matches that
  correspond to structures added during integration. The lower panel identifies
  the structures and property data used for screening and novelty assessment.
  (b) Property values contributed by each database. Sources are numbered in
  descending order of property-value count, starting at 1, with names in
  Supplementary Table~\ref{si-tab:db-numbers}.
  (c) The 45 databases containing at least 1{,}200 crystals, coloured as in (b).
  (d) Source references by database family and matching method.
  (e) The 36 databases with at least 1{,}000 matches. Marker area represents the
  number of matches, and colour represents the percentage that correspond to
  structures added during integration.
  (f) Crystals recorded in one database (grey) or two or more databases (blue).}
  \label{fig:record}
  \alttext{(a) A workflow links six database entries through four matching and
  integration steps, with a dashed loop showing reference-set updates and a panel
  identifying the data used by the two design paradigms. (b) Bars show property
  values per database on a logarithmic scale. (c) The scatter plot compares the
  number of structures in each database with the percentage absent from other
  sources. (d) Stacked bars show source references by family and matching method.
  (e) Points show the percentage of matches to newly added structures in ingestion
  order. (f) A logarithmic histogram shows the number of databases per crystal.}
\end{figure*}

Broader database coverage expands the structures available for forward screening
and helps identify generated crystals that are already known. No single database contains all known crystals, so we compiled a comprehensive collection of 55 crystal databases, spanning major
general-purpose computational repositories, experimental structure databases
and specialized materials collections. Each source contributes entries
for individual crystals, with their available property values, spectra and source
information. We match
each incoming entry to the structures already assembled, first using an
authoritative cross-database identifier published by its source and then, for
unresolved entries, a structural fingerprint. Chemical formula alone cannot
distinguish polymorphs (Fig.~\ref{fig:record}a and Section~\ref{sec:methods};
Supplementary Note~\ref{si-note:fingerprint}). For example,
the CaMg$_2$Sb$_2$ structure in Fig.~\ref{fig:record}a has different identifiers
across 17 of the 55 databases. Matching these entries links 21 source entries
to one crystal.
An unmatched entry is added as a new structure, and every property value retains
its source database, native entry and calculation or measurement method.

Applied to all 55 sources, this matching yields the 14{,}590{,}485 distinct crystal structures of MateriaViva, 357{,}127
from experimental structure databases and 14{,}233{,}358 computational candidates, together with
131{,}824{,}881 property values for 166 properties, 1{,}198{,}897 stored spectra
and 21{,}543{,}633 links to source entries. Experimental-source provenance is distinguished from synthesis evidence,
which is assessed in Section~\ref{sec:screening}.
The 55 sources contribute very unequally to these totals, spanning 5.7 orders of magnitude. The
largest supplies 46.0\% of all values, whereas 30 of the 55 supply fewer than $10^5$
each (Fig.~\ref{fig:record}b). Each database also contains a different subset of known
crystals (Fig.~\ref{fig:record}c). Among the 5{,}894{,}222
crystals in Alexandria \cite{schmidt2022dataset}, 75.4\% are absent from every
other integrated database, compared with only 1.3\% of the
211{,}854 crystals in the Materials Project \cite{jain2013commentary}. The
share is 100\% for the enumerated framework libraries, whose crystals are absent
from all other integrated databases, and below 1\% for several specialized libraries that
contain almost exclusively crystals already recorded in other databases. A screening study that
uses one database therefore excludes structures available only in the other
databases.

Identifying crystals shared across multiple databases requires explicit structure matching.
Among the links to source entries, 6{,}784{,}564 were resolved through a published
cross-database identifier and 8{,}294{,}560 through a matching structural
fingerprint. The remaining links point directly to each crystal's entry in its source
database, including sources that provide properties only
(Fig.~\ref{fig:record}d). Because each newly admitted structure joins the matching set for all subsequent
sources, 3{,}676{,}707 of the 15{,}079{,}124 links resolved by matching (24.4\%)
refer to structures added during the integration itself (Fig.~\ref{fig:record}e).
Expanding the matching set in this way prevents crystals shared by later sources
from being counted again as new structures. NOMAD, ingested last, resolved 56.1\% of its
3{,}168{,}530 matches onto structures added during integration and contributed
1{,}047{,}208 structures of its own (Supplementary Note~\ref{si-note:order}). Likewise, assessing generated crystals
against a limited set of crystal databases can overestimate novelty
(Section~\ref{sec:novelty}). Most of the 13{,}902{,}836 structures linked
to at least one source entry occur in only one database, but 3{,}233{,}769
occur in two or more and a few hundred in twenty or more
(Supplementary Table~\ref{si-tab:sources})
(Fig.~\ref{fig:record}f). These shared structures make it possible to compare
property values reported by different databases.

%% file: results_labels.tex
\phantomsection\label{sec:inherits}

\begin{figure*}
  {\centering\includegraphics{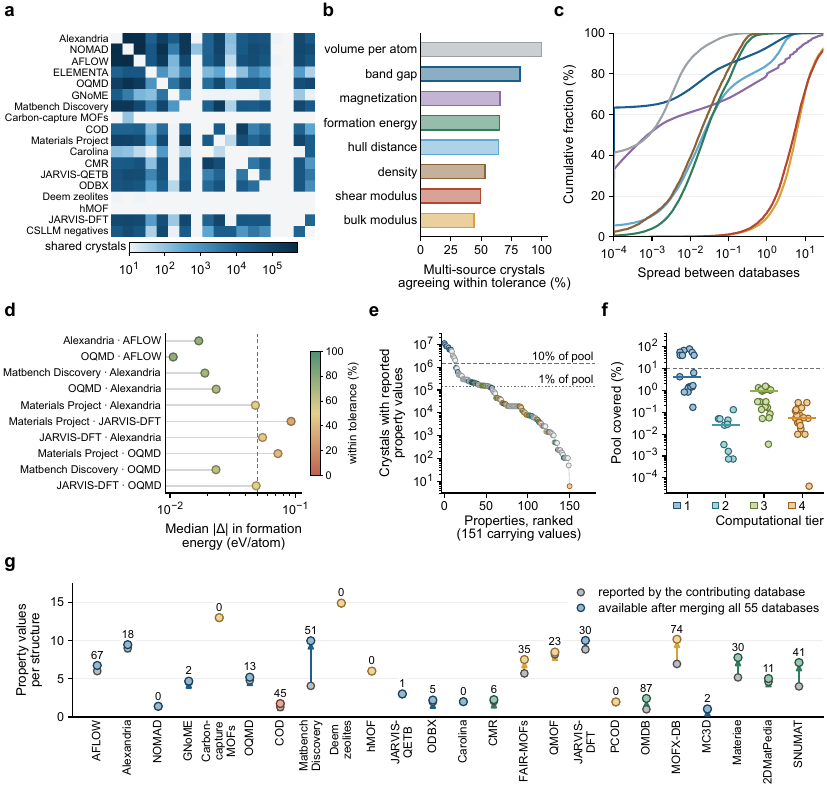}\par}
  \caption{\textbf{Cross-database agreement and property coverage.}
  (a) Crystals shared between the 18 databases containing the most structures.
  Empty squares indicate pairs with no shared structures.
  (b) Fraction of multi-source crystals with property values agreeing within the
  tolerances in Supplementary Note~\ref{si-note:agreement}.
  (c) Cross-source spread in each property's units, expressed as a relative
  difference for density and volume per atom, with colours as in (b).
  (d) Formation-energy differences for the ten database pairs sharing the most
  crystals. The dashed line marks the agreement tolerance.
  (e) Property coverage, coloured by the tiers in (f), with unassigned properties
  in grey.
  (f) Coverage of general-purpose properties by calculation tier: one
  self-consistent calculation~(1), additional self-consistent calculations~(2),
  linear response~(3), and anharmonic, transport or beyond-semilocal
  calculations~(4). Points represent properties, bars are medians, and the dashed
  line marks 10\% coverage.
  (g) Mean values per structure for the structures introduced by each of the
  24 largest contributors. The percentage above each source gives the fraction
  of its structures that gained a property value absent from that source.
  Colours identify database families (Fig.~\ref{fig:record}b).}

  \label{fig:problem}
\alttext{(a) A heat map shows the number of crystals shared by 18
  databases. (b) Bars show the fraction agreeing within tolerance for eight
  properties. (c) Cumulative curves show cross-database property spread.
  (d) Points show median formation-energy differences for ten database pairs.
  (e) Property coverage is plotted by rank. (f) Points show coverage in four
  calculation tiers. (g) Paired bars compare values per structure before and
  after integration.}

\end{figure*}

Because forward screening relies on recorded property values to train predictors
and rank candidates, we first evaluated the consistency of these values across all
55 sources. Shared structures are concentrated in a few database pairs, with the
largest overlap containing 1{,}672{,}167 shared crystals (Fig.~\ref{fig:problem}a). The
property values for these shared structures can be compared because every value
retains the method that produced it. We map each method description to a level of theory and compare two
databases only when they report the same property at a comparable level
(Fig.~\ref{fig:problem}b--d; Supplementary Note~\ref{si-note:agreement} and
Supplementary Table~\ref{si-tab:pairs}).
Among the 151 properties with reported values (Supplementary
Table~\ref{si-tab:properties}), 66 are reported by two or more
databases and can be compared in this way, whereas the other 85 have a single supplier.
Agreement varies across properties under the property-specific tolerances. Within 50~meV/atom,
65.3\% of formation energies and 64.2\% of hull distances agree, compared with
82.1\% of band gaps within 50~meV and 44.3\% of bulk moduli within 5~GPa. The largest disagreements can change stability classifications: the
worst-agreeing tenth of shared formation energies differ by more than 205~meV/atom (Fig.~\ref{fig:problem}c),
several times the hull-distance margins conventionally used to call a compound
stable \cite{sun2016thermodynamic,bartel2020critical}. Whether a candidate passes
such a filter can therefore depend on which database supplied its energy, and
measuring this dependence requires comparing energies for the same structure
across databases.
The retained provenance also shows that some of the most closely agreeing pairs
report one underlying calculation that two databases have both ingested, which
distinguishes independent confirmation from repeated data (Supplementary Note~\ref{si-note:agreement-reading}).

Beyond the consistency of recorded values, their availability constrains screening: although 94.5\% of the crystals have at least one reported property value,
coverage of any particular property is usually low. Across the 151 of the 166
properties with reported values, only 4.52\% of crystal--property pairs have a
value
(Fig.~\ref{fig:problem}e; Supplementary Note~\ref{si-note:occupancy} and
Supplementary Fig.~\ref{si-fig:contents}). Occupancy
is uneven, with nine properties covering more than 40\% of the structures and 97 of the
151 covering less than 1\%. Coverage initially tracks computational cost but then breaks from this trend
(Fig.~\ref{fig:problem}f; Supplementary Note~\ref{si-note:tiers} and
Supplementary Table~\ref{si-tab:tiers}). Among the 69
general-purpose properties, the 19 obtained from a standard self-consistent
calculation have a median coverage of 4.23\% of database structures, whereas
the 50 that require a dedicated calculation have a median of 0.0995\%, 43
times lower, and none of them reaches 10\%. Within this second group, however,
the more expensive linear-response properties are better covered
(median 0.987\%) than properties obtainable from a few extra self-consistent runs
(median 0.0269\%), a 37-fold reversal of the expected cost trend. The elastic and dielectric tensors
were the targets of dedicated high-throughput campaigns
\cite{dejong2015charting,petousis2017high}, whereas less expensive quantities
such as the work function were not covered by comparable campaigns in the integrated sources. Coverage therefore reflects the history of dedicated high-throughput studies
as well as computational cost.

Even this 4.52\% coverage already includes the values gained by
linking the same structure across databases (Fig.~\ref{fig:problem}g). Among the 6{,}266{,}517 structures
added during integration that have any property value, 26.1\% (1{,}637{,}460)
gain at least one value absent from the database that introduced them.
Across the whole database, 22.9\% of the crystals with any property value
(3{,}153{,}748) draw values from more than one database, a share that does not
depend on the order in which the sources are ingested. Building on this
order-invariant incremental matching, a background autonomous agent continuously
detects newly released crystal repositories and integrates them into MateriaViva
so that the database self-expands over time. Integration thus expands both the
structures available for screening and the property data available to train and
evaluate prediction models.

%% file: results_completion.tex
\subsection{Property completion with an equivariant graph transformer}
\label{sec:model}

\begin{figure*}
  {\centering\includegraphics{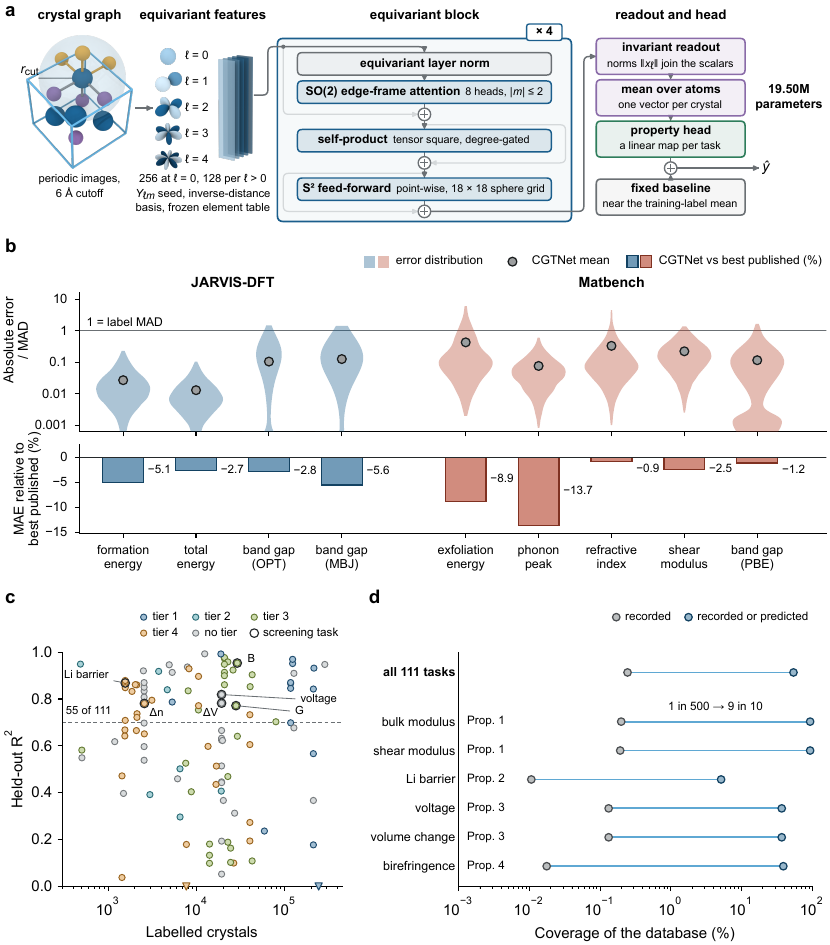}\par}
\caption{\textbf{CGTNet architecture, benchmark performance and property completion.}
  (a) The architecture maps a periodic crystal graph to a task-specific scalar
  prediction. Each task uses a separately trained model. Grey brackets indicate
  residual connections, and the block multiplier denotes four stacked blocks.
  The CGTNet output is added to a fixed baseline close to the
  mean of the training labels. The configuration is given in Supplementary
  Table~\ref{si-tab:gnn-config}.
  (b) Errors for four JARVIS-DFT properties and five Matbench tasks
  (Table~\ref{tab:bench}). The upper panel shows absolute errors divided by the
  mean absolute deviation (MAD) of the target labels. Violins give the
  distribution over evaluated crystals, with JARVIS-DFT in blue and Matbench in
  red, and points show CGTNet's mean error. The lower panel gives the percentage
  difference between CGTNet's error and the lowest error among the other methods
  in Table~\ref{tab:bench}. Negative values indicate a lower error for CGTNet.
  (c) Held-out $R^2$ versus the number of labelled crystals for the 140
  models, coloured by calculation tier (Fig.~\ref{fig:problem}f). Rings mark
  screening tasks, the dashed line marks $R^2 = 0.7$, and triangles on the
  horizontal axis mark tasks with $R^2$ below zero.
  (d) Fraction of database structures with a recorded value (grey) or a recorded
  or predicted value (blue), aggregated over completed tasks and shown separately
  for each screening task.}

  \label{fig:model}
\alttext{(a) The model diagram shows a periodic radius graph, four equivariant
  blocks and a scalar output added to a fixed baseline. Each task
  has an independently trained model. (b) Violins show absolute errors divided
  by target-label MAD for four JARVIS-DFT and five Matbench tasks. Points mark
  mean errors, and bars compare CGTNet with the lowest reported error among the
  other listed methods.
  (c) Held-out R squared is plotted against the number of labelled structures.
  Of the models shown, 74 reach 0.7. (d) Paired points compare recorded-property
  coverage with coverage after prediction, overall and for the screening tasks.}

\end{figure*}

Integration removes duplicate structures and unifies the property values
that the 55 sources report, yet more than 95\% of crystal--property pairs remain
empty afterwards because the original studies measured or calculated only
selected properties. Property completion predicts these missing values, so that
each structure can be screened for properties beyond its original record.
The completion library comprises 140 property-prediction models
applied across the database of 14{,}590{,}485 crystals (Supplementary
Note~\ref{si-note:census}). Accurate completion at this scale requires learning from label sets of widely
varying sizes and predicting values for millions of unlabelled crystals.
We therefore built one architecture and one training recipe for
completion tasks, with property-specific neighbour graphs and target
transformations. CGTNet (Fig.~\ref{fig:model}a) is a rotation-equivariant graph transformer that
combines the design principles of two model families (Supplementary
Note~\ref{si-note:model} and Supplementary Table~\ref{si-tab:gnn-config}). From equivariant interatomic potentials, it adopts spherical-harmonic channels up
to degree four mixed within the frame of each edge; from invariant property models,
it incorporates an inverse-distance radial coordinate, a fixed physical element
table and summed neighbour messages. The angular channels of each atom are additionally
initialized from the directions of its own neighbours, so that directional
information about the local geometry is available before the first message is
passed.

Before applying CGTNet to the database, we tested it on two public benchmarks
(Table~\ref{tab:bench}; Fig.~\ref{fig:model}b; Supplementary
Tables~\ref{si-tab:gnn-jarvis} and~\ref{si-tab:gnn-matbench};
Supplementary Fig.~\ref{si-fig:parity}). On JARVIS-DFT \cite{choudhary2020jarvis}, using the benchmark settings described in Section~\ref{sec:methods}, its error on each of the four properties is
below the lowest of at least \MinJarvisEntries{} published results for that
property. On Matbench \cite{dunn2020benchmarking}, CGTNet has the lowest five-fold mean
among the methods compared on the exfoliation energy, the phonon peak frequency,
the refractive index, the logarithmic shear modulus and the PBE band gap
(Supplementary Note~\ref{si-note:model}).
Both benchmarks split crystals at random. For database completion, we apply the
stricter test of held-out chemical systems, which keeps compounds with the same constituent
elements in one split and tests predictions for element combinations absent from
training.

\begin{table*}
\caption{\textbf{Benchmark errors of CGTNet and published methods.} Mean absolute
errors on the standard test split of four JARVIS-DFT properties and over the five
official folds of five Matbench tasks: exfoliation energy (jdft2d), the frequency
of the highest phonon density-of-states peak (phonons), refractive index
(dielectric), logarithmic shear modulus (log\,$G$) and PBE band gap (mp\_gap).
Bold indicates the lowest value in each column, and a dash indicates that a
method does not report the task. Settings and the sources of the published
values are given in Section~\ref{sec:methods}.}

\label{tab:bench}
\tablefont
\begin{tabular*}{\textwidth}{@{\extracolsep\fill}lrrrrrrrrr@{\extracolsep\fill}}
\toprule
 & \multicolumn{4}{c}{JARVIS-DFT} & \multicolumn{5}{c}{Matbench} \\
\cmidrule(lr){2-5}\cmidrule(lr){6-10}
Method & $E_\mathrm{f}$ & $E_\mathrm{tot}$ & $E_\mathrm{g}^\mathrm{OPT}$ & $E_\mathrm{g}^\mathrm{MBJ}$ & jdft2d & phonons & dielectric & log\,$G$ & mp\_gap \\
 & meV/atom & meV/atom & meV & meV & meV/atom & cm$^{-1}$ & -- & -- & meV \\
\midrule
CGCNN & 62.5 & 81.5 & 190.8 & 406.7 & 49.2 & 57.8 & 0.5988 & 0.0895 & 297.2 \\
SchNet & 45.0 & 47.0 & 190.0 & 430.0 & 42.7 & 39.0 & 0.3277 & 0.0796 & 235.2 \\
MEGNET & 47.0 & 58.0 & 145.0 & 340.0 & 54.2 & 28.8 & 0.3391 & 0.0871 & 193.4 \\
ALIGNN & 33.1 & 36.7 & 142.3 & 310.4 & 43.4 & 29.5 & 0.3449 & 0.0715 & 186.1 \\
Matformer & 32.2 & 35.0 & 137.0 & 300.0 & -- & -- & -- & -- & -- \\
PotNet & 29.3 & 32.1 & 124.6 & 270.7 & -- & -- & -- & -- & -- \\
eComFormer & 28.4 & 32.0 & 124.0 & 280.0 & -- & -- & -- & -- & -- \\
iComFormer & 27.2 & 28.8 & 122.0 & 260.0 & -- & -- & -- & -- & -- \\
CartNet & 27.1 & 26.6 & 115.3 & 253.0 & -- & -- & -- & -- & -- \\
PRISM & 25.9 & 26.3 & 109.3 & 236.5 & -- & -- & -- & -- & -- \\
PRDNet & 25.0 & 24.0 & -- & 242.0 & -- & -- & -- & -- & -- \\
ReciNet & 27.0 & 26.8 & 126.0 & 240.0 & 31.6 & -- & -- & -- & -- \\
MODNet & -- & -- & -- & -- & 33.2 & 34.3 & 0.2711 & 0.0731 & 219.9 \\
coGN & -- & -- & -- & -- & 37.2 & 29.7 & 0.3088 & 0.0689 & 155.9 \\
coNGN & -- & -- & -- & -- & 36.2 & 28.9 & 0.3142 & 0.0670 & 169.7 \\
\midrule
\textbf{CGTNet} & \textbf{23.7} & \textbf{23.4} & \textbf{106.2} & \textbf{223.3} & \textbf{28.8} & \textbf{24.8} & \textbf{0.2687} & \textbf{0.0653} & \textbf{154.1} \\
\bottomrule%
\end{tabular*}
\end{table*}

\phantomsection\label{sec:completion}
For completion across the integrated database, each task predicts one property at a specified level of theory and,
where relevant, under specified conditions. Splitting the 151 populated
properties by level of theory yields 513
candidate tasks (Supplementary Fig.~\ref{si-fig:completion-sets}a), whereas
existing property-prediction and stability benchmarks provide tens
\cite{dunn2020benchmarking,riebesell2025framework}. For each property we kept
the level of theory with the most labels and split condition-dependent
properties by their conditions. Property completion covers 140 of the resulting
properties, each with its own model (Supplementary Note~\ref{si-note:census}). Different levels of theory are treated as separate
tasks because the systematic offsets between them, of order 1~eV for the band gap,
cannot be inferred from the crystal structure alone (Supplementary
Fig.~\ref{si-fig:theory}). Finally, each task is restricted to an explicit
applicability domain of structures for which its property is physically defined
(Supplementary Fig.~\ref{si-fig:completion-sets}b).

The 140 models add 874{,}091{,}330 predicted property values
(Supplementary Notes~\ref{si-note:model} and~\ref{si-note:completion};
Supplementary Table~\ref{si-tab:completion-tasks}),
which provide at least one prediction for 14{,}567{,}993 crystals, or
99.85\% of the database. On held-out chemical systems, 74 models reach a coefficient of determination
($R^2$) of at least 0.7, up from 44 at the pre-registration snapshot
(Fig.~\ref{fig:model}c; Supplementary Fig.~\ref{si-fig:task-accuracy}),
and the application examples and the equal-budget comparison below use only tasks
finished by that snapshot.
For the semilocal bulk modulus, completion raises coverage from 0.20\% of
structures with recorded values to 93.2\% with recorded or predicted values
(Fig.~\ref{fig:model}d). This allows screening of known crystals whose bulk
moduli are absent from every integrated database. Each completed task is released with its trained model, applicability domain and
performance estimates on held-out chemical systems as part of an open
property-prediction model library on Hugging Face, so generative studies can use
the same models to predict properties of their outputs within the stated domains,
as the bulk-modulus comparison below demonstrates.

We further tested the predictions with DFT directly on the unlabelled structures
that completion targets. For a random sample of database structures without a
recorded bulk modulus (S-all), the mean absolute error of the bulk-modulus
predictions on the \dftSallN{} structures evaluated
is \dftSallmae~GPa, compared with 7.4~GPa on the holdout. DFT also verifies
selected mechanical and optical candidates and the candidates of the equal-budget
comparison (Sections~\ref{sec:screening} and~\ref{sec:methods}). Compared with the
spread between sources, the formation-energy benchmark error is
\FormationMeV~meV/atom and lies within the 50~meV/atom agreement tolerance for
\WithinAgreeWindow{}\% of held-out crystals. Across the 88 tasks
with non-zero median cross-source spread, twelve models already reach held-out mean
absolute errors within that spread (76 remain above it), meaning that for these
twelve properties predictions on unseen chemical systems deviate from the reference
labels no more than independent databases deviate from one another (Supplementary
Fig.~\ref{si-fig:parity} and Supplementary Note~\ref{si-note:model}).

\subsection{Limits and coverage of forward screening}
\label{sec:screening}

\begin{figure*}
  {\centering\includegraphics{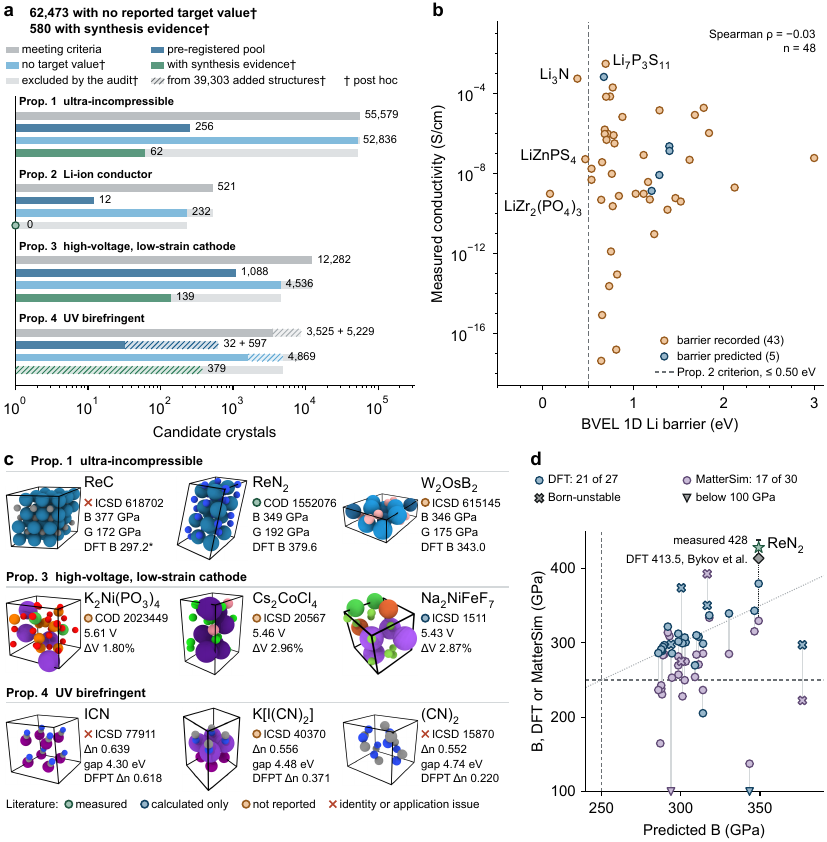}\par}
\caption{\textbf{Four application examples of property completion and candidate verification.}
  (a) Candidate counts for each target after the pre-registered criteria,
  restriction to the pre-registered subset, the post hoc audit for absence of
  recorded primary-property values, and restriction to synthesis evidence.
  Daggers mark the latter two post hoc categories. Light segments show candidates
  removed by the audit. Hatching marks Prop.~4 candidates from the
  structures added to its applicability domain during the audit.
  (b) Recorded (orange) or predicted (blue) Prop.~2 barriers versus measured
  room-temperature Li-ion conductivity for 48 crystals with both quantities.
  Values are medians across each crystal's entries. The dashed line marks the
  barrier criterion, and the band-gap criterion is not applied.
  (c) Three highest-ranked candidates with synthesis evidence for each target
  after the post hoc audit, with none remaining for Prop.~2. The panels give
  predicted values and, for Prop.~4, recorded gaps. Held-out model errors are
  listed in Supplementary Table~\ref{si-tab:completion-tasks}. Prop.~1 panels
  also report the PBE bulk modulus in GPa (DFT $B$), with an asterisk marking a
  Born-unstable tensor. Prop.~4 panels report PBE birefringence (DFPT $\Delta n$).
  Symbols identify the literature status of the screened property.
  (d) Predicted bulk moduli of the 30 highest-ranked audited Prop.~1 candidates
  versus PBE (blue) and MatterSim (purple) values. Crosses mark Born-unstable
  tensors, triangles indicate values below 100~GPa, and dashed lines mark
  250~GPa. Counts require a bulk modulus of at least 250~GPa and a mechanically
  stable elastic tensor.
  For ReN$_2$, the star marks the measurement and the diamond the DFT value
  reported in the same experimental study \cite{bykov2019rhenium}.}

  \label{fig:screening}
\alttext{(a) Horizontal bars on a logarithmic scale show candidate counts
  through the original screening criteria, pre-registered subset, post hoc
  audit and synthesis-evidence restriction. (b) A scatter plot compares measured
  Li-ion conductivity with recorded or predicted migration barriers.
  (c) Nine crystal structures are shown with predicted properties and symbols
  indicating the literature evidence. (d) DFT and MatterSim bulk moduli are
  plotted against model predictions for 30 selected crystals.}

\end{figure*}

With these predictions, known crystals can be screened for properties that no
integrated database records for them. For each property covered by the
completed tasks, we screened the crystals within its applicability domain. We present four application examples, Property~1 to Property~4
(Prop.~1--4). Their thresholds were pre-registered before any candidate was counted,
and the examples use only tasks with $R^2 \ge 0.7$ (Supplementary Note~\ref{si-note:prereg}). Prop.~1 seeks ultra-incompressible solids
(bulk modulus $\ge 250$~GPa and shear modulus $\ge 150$~GPa) and Prop.~2 Li-ion
conductors (one-dimensional bond-valence Li migration barrier $\le 0.50$~eV and
recorded semilocal gap $\ge 3.0$~eV). Prop.~3 seeks high-voltage, low-strain cathodes
(voltage $\ge 4.0$~V and volume change $\le 3\%$) and Prop.~4 UV birefringent
crystals (birefringence $\ge 0.10$ and recorded semilocal gap $\ge 4.0$~eV). The
screens return 55{,}579, 521, 12{,}282 and 3{,}525 candidates
(Fig.~\ref{fig:screening}a; Supplementary Table~\ref{si-tab:screening-counts}). Among the Prop.~1 candidates, 54{,}953 pass on predicted
values alone and 53{,}813 have no recorded bulk modulus. The pre-registered
subsets, which contain crystals from experimental structure databases
\cite{zagorac2019recent,grazulis2009crystallography} lacking a recorded primary-property value in any
source, comprise 256, 12, 1{,}088 and 32 candidates (Supplementary
Tables~\ref{si-tab:screening-attrition} and~\ref{si-tab:showcase}).

Comparing the highest-ranked candidates of each pre-registered subset with
the literature revealed four pitfalls that database records introduce into a
conventional screen (Supplementary Note~\ref{si-note:screening}). The
ultra-incompressible (Prop.~1) selections were
theoretical structures catalogued in an experimental structure database
\cite{zagorac2019recent}, and the Li-ion conductor (Prop.~2) selections were crystals whose barrier
is already recorded under another entry, among them the known Li-ion conductors
LiB$_3$O$_5$ and $\beta$-eucryptite
\cite{guo1995pyroelectric,vonalpen1977eucryptite}. The cathode (Prop.~3) selections were fluorides
without a redox-active cation, and the UV birefringent (Prop.~4) selections were coordination polymers from a
material family absent from the training set. We therefore used cross-database provenance to establish a post hoc audited
screen, fixed before any audited candidate was counted, with one rule for each
pitfall. It excludes candidates when another entry with the same reduced formula and
space group records the primary-property value, or when their elements or
material family lack training-set coverage. Prop.~3 also requires a redox-active
cation. Synthesis evidence is assessed separately within the audited set. We define
synthesis evidence as a recorded R factor, recorded diffraction conditions or a
Materials Project match to an experimental entry, criteria met by 294{,}687 of the
357{,}127 structures from experimental structure databases. The audit also extended birefringence prediction to 39{,}303 structures excluded
from the Prop.~4 applicability domain by placeholder band-gap values. The audited screen leaves 52{,}836, 232, 4{,}536 and
4{,}869 candidates without a recorded primary-property value, of which 62, 0, 139 and 379 have confirmed
synthesis evidence.

Property completion identifies synthesized crystals without a recorded
target-property value in three of the four audited applications (Fig.~\ref{fig:screening}c). DFT confirms
the mechanical and optical criteria for selected candidates. Among the \dftTonesynthN{} synthesized Prop.~1 candidates evaluated by
DFT, \dftTonesynth{} meet both modulus thresholds with mechanically stable
elastic tensors, and \dftTonesynthB{} meet the bulk-modulus and mechanical-stability
criteria. Setting the shear criterion aside, \dftTonepre{} of the
\dftTonepredone{} pre-registered Prop.~1 selections and \dftTonetop{} of the
\dftTonetopdone{} highest-ranked audited candidates with synthesis evidence satisfy
both the bulk-modulus threshold of 250~GPa and Born elastic stability
\cite{mouhat2014necessary} in DFT, compared with
\msTonepre{} of 30 and 17 of 30 under MatterSim \cite{yang2024mattersim}. The shear criterion is also met by \dftTonetopG{} of the DFT-confirmed audited candidates
(Fig.~\ref{fig:screening}d; Supplementary Notes~\ref{si-note:screening}
and~\ref{si-note:mlip}).
All synthesized UV birefringent (Prop.~4) candidates come from
structures added to its domain during the audit. Density-functional perturbation
theory (DFPT) gives a birefringence of at least 0.10 for \dftTfourbiref{} of the
\dftTfourN{} evaluated candidates, and \dftTfourpass{} of these also have a
Perdew--Burke--Ernzerhof (PBE) gap of at least 4.0~eV. In the pre-registered
subset, \dftTfourprebiref{} of \dftTfourpreN{} evaluated structures reach the
birefringence threshold, and \dftTfourprepass{} also meet the gap criterion.

For the Li-ion conductors (Prop.~2), the audit traces the absence of synthesized candidates to the
databases themselves. All 12 crystals in its pre-registered subset share their
formula and space group with entries that have a recorded barrier, so they are
already characterized, and none of the 232 audited candidates comes from an
experimental structure database. Across the 48 crystals with both quantities, the
bond-valence migration barrier shows no rank correlation with measured
room-temperature conductivity (Spearman $\rho = -0.03$; Fig.~\ref{fig:screening}b).
The integrated records thus locate the limit of this screen in the descriptor
available for Li-ion transport, whereas the other three applications return synthesized candidates.

Among the audited Prop.~1 candidates with synthesis evidence, ReN$_2$ was synthesized at high
pressure and has a measured bulk modulus of 428~GPa and a hardness of 36.7~GPa
\cite{bykov2019rhenium}. None of the 55 integrated databases reports a bulk modulus for this
structure, so property completion recovered a known ultra-incompressible solid.
CGTNet predicts 349~GPa, well above the 250~GPa criterion, and PBE gives \dftReNtwo~GPa here and 413.5~GPa in the study that
reported the measurement \cite{bykov2019rhenium}. For the synthesized crystals W$_2$OsB$_2$, K$_2$Ni(PO$_3$)$_4$ and
Cs$_2$CoCl$_4$ we found no report of the screened property, so these predictions
point to properties that have yet to be measured. Property completion thus identifies high-performance candidates among known
crystals whose target-property values are absent from the integrated databases.
After the provenance audit, 580 of these candidates have synthesis evidence. The potential of forward screening is therefore underestimated when
it is judged only from the property values currently reported in databases.

%% file: results_novelty.tex
\subsection{Limits and coverage of inverse design}
\label{sec:novelty}

The database that supplies candidates for screening is also the reference against
which generated crystals are tested for novelty. The assessments cited in the Introduction
\cite{cheetham2024artificial,leeman2024challenges,juelsholt2026continued,szymanski2025baselines}
decided matches in different ways, ranging from StructureMatcher tolerances
against ordered entries to expert comparison with disordered phases, so the same
crystal can be new under one criterion and known under another. A criterion may,
for example, count a known prototype with different occupants as new, or miss an
ordered approximant of a parent phase that is disordered on one site (Supplementary Note~\ref{si-note:novelty}).

\begin{figure*}
  {\centering\includegraphics{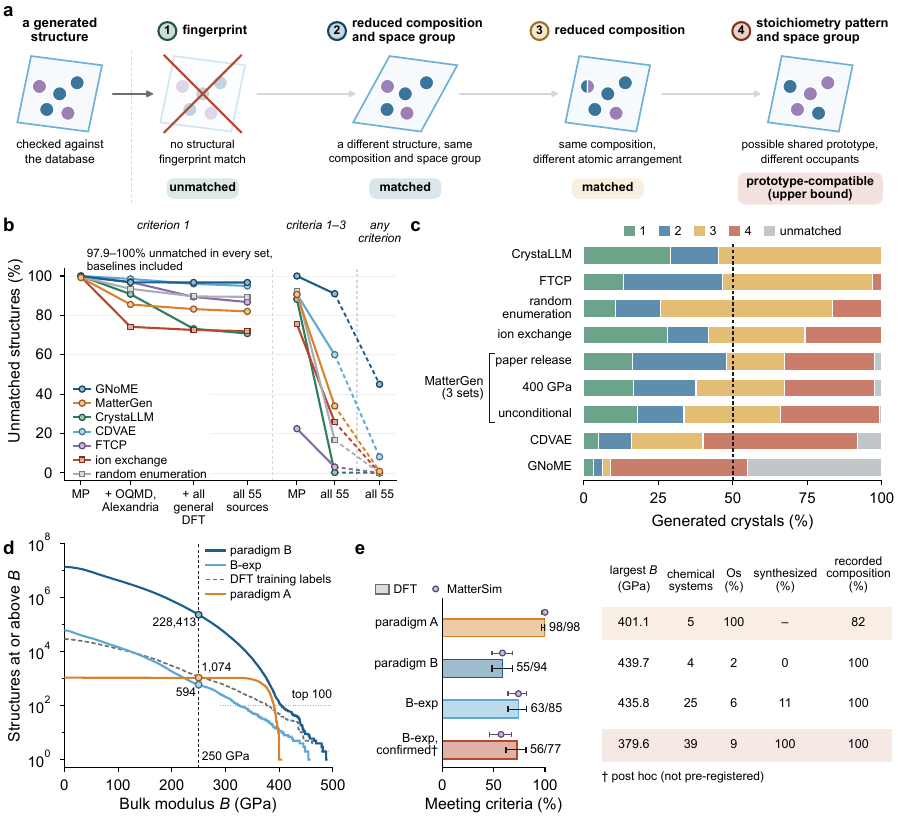}\par}
\caption{\textbf{Novelty assessment and comparison of generation with screening.}
  (a) A generated crystal evaluated under four crystal-matching criteria. Criterion 3
  identifies a shared composition even when atomic arrangements differ.
  Criterion 4 is necessary but insufficient for a shared prototype, so its
  matches give an upper bound on possible prototype redundancy.
  (b) Fractions without a match under criterion 1 as the crystal databases used for comparison expand from
  Materials Project (MP) to all 55 sources. Further comparisons apply criteria
  1--3 to MP and all 55 sources, and all four criteria to all 55 sources (dashed).
  MatterGen denotes the unconditional run, and squares mark the two baselines.
  (c) The nine sets grouped by the strictest criterion matching each crystal
  against all 55 sources, ordered by the fraction matched under criteria 1--3.
  Grey marks structures unmatched under any criterion. The dashed line marks
  half of each set.
  (d) Numbers of structures with predicted bulk modulus at or above $B$ for
  inverse design (paradigm A, MatterGen conditioned on 400~GPa), forward
  screening (paradigm B, database structures without recorded bulk moduli) and
  B-exp (the experimental-source subset of paradigm B),
  together with the DFT labels of the bulk-modulus task. Lines mark the
  bulk-modulus threshold and verification budget.
  (e) Fractions satisfying $B \geq 250$~GPa and mechanical stability in DFT
  (bars) and MatterSim (purple circles), with Wilson 95\% intervals. MatterSim
  evaluates each full list of 100 structures, and labels give the number
  satisfying the criteria over the number evaluated. The synthesis-evidence subset of B-exp is marked with
  a dagger as post hoc. The table reports, for DFT-confirmed candidates, the
  maximum $B$, number of chemical systems and fractions containing Os, with
  synthesis evidence and with recorded compositions (criteria 1--3). Shading
  identifies paradigm A and the synthesis-evidence subset. Colours in (d,e)
  identify the candidate sets.}

  \label{fig:novelty}
\alttext{(a) A schematic compares a generated crystal under four crystal-matching
  criteria. (b) Points show fractions of seven structure sets without matches
  as more crystal databases are included and matching criteria become more permissive.
  (c) Stacked bars divide all nine sets by matching criterion, with a line at
  half of each set. (d) Logarithmic curves show structure counts above each
  predicted bulk-modulus threshold. (e) Bars and points compare DFT and
  MatterSim fractions meeting the bulk-modulus and mechanical-stability
  criteria. A table describes the DFT-confirmed candidates.}

\end{figure*}

To quantify these effects, we audited nine structure sets: GNoME's contribution to this database
\cite{merchant2023scaling}, three MatterGen sets (two of them generated from
released checkpoints \cite{zeni2025generative}) and the 500-structure samples
released for CDVAE \cite{xie2022crystal}, CrystaLLM \cite{antunes2024crystal},
FTCP \cite{ren2022ftcp}, ion exchange and random enumeration
\cite{szymanski2025baselines}. Against all 55 sources, four
criteria are applied in turn from strict to permissive (Fig.~\ref{fig:novelty}a),
matching the crystal itself by structural fingerprint, the reduced composition
with the space group, the reduced composition alone, and the stoichiometry
pattern with the space group. Under fingerprint matching against Materials Project alone, the crystal database commonly used for novelty assessment, 97.9--100\%
of every audited set remains unmatched, including the two baselines.
This fraction of generated structures absent from Materials Project therefore cannot
distinguish the audited generators from random enumeration. Against all
55 sources, between 3.3\% (GNoME) and 29.2\% (CrystaLLM) of each set is matched by the structural fingerprint (Fig.~\ref{fig:novelty}b). Including all 55 crystal databases changes the match counts more under criteria 1--3. Materials
Project matches 0.0--24.4\% of each set, with FTCP the exception at 77.6\%,
whereas all 55 sources match 9.0--99.8\%. CrystaLLM's share with a recorded
composition, for instance, rises from 11.8\% to 99.8\%. The wider database therefore identifies shared compositions missed by Materials Project. The most stringent novelty test requires a structure to remain unmatched under
all four criteria. Under this test, 55--100\% of each set fails to qualify as
new, leaving 0.0--8.2\% of the small sample sets and 45.0\% of GNoME unmatched
(Fig.~\ref{fig:novelty}a,c; Supplementary Note~\ref{si-note:fingerprint} and
Supplementary Fig.~\ref{si-fig:novelty-holders}).

Claims of compositional novelty should therefore use the third criterion,
which reports the share of structures whose reduced composition is absent from
the database, separately from the novelty of the atomic arrangement. This share is 91.0\% for GNoME,
60.0\% for CDVAE, 34.0\% for MatterGen, 25.8\% for ion exchange, 16.7\% for
random enumeration, 3.0\% for FTCP and 0.2\% for CrystaLLM. For the three MatterGen runs, 66.0--67.4\% of structures have a recorded composition.
GNoME also contains more complex compositions: 87.4\% of its released set
contains four or five elements, compared with only 29.2\% of the database.
Because such compositions are sparsely represented in the
crystal databases used for comparison, their absence does not by itself establish
that the compounds have never been reported. Accounting for disordered phases
lowers these shares further, because partially occupied entries are excluded
from fingerprinting and the composition-level shares are therefore upper bounds
on novelty. Reading compositions from such
entries changes the nine sets by at most 0.4 percentage points, and against the
experimental entries it reduces the A-Lab targets \cite{szymanski2023autonomous}
with an unrecorded composition from 57 to 36 of 58
(Supplementary Note~\ref{si-note:disorder}).

By contrast, the CrystaLLM and random-enumeration samples both consist entirely of ternary
compounds, allowing a comparison within the same composition complexity.
Under fingerprint matching, 70.8\% and 89.4\% of their structures, respectively,
are absent from the database, a difference of 18.6 percentage points. At the composition level the two differ far more: 499 of the 500 CrystaLLM
structures (99.8\%) have a composition already recorded in the database, compared
with 410 of 492 (83.3\%) for random enumeration, so their fractions of unrecorded
compositions (0.2\% versus 16.7\%) are more than eightyfold apart.
CrystaLLM thus largely reproduces recorded compositions, whereas even random
enumeration produces compositions absent from the integrated database. The fraction of generated crystals
classified as new thus depends on the crystal databases and matching criteria as
well as on the generator.

\phantomsection\label{sec:route}

The novelty audit asks whether generated crystals are new, but a complementary
question is whether generation finds better candidates than screening.
MatterGen's own evaluation found 106 stable, unique and new structures with a bulk
modulus above 400~GPa within 180 DFT property calculations, against 40 for a screening
baseline that used the same property predictor \cite{zeni2025generative}. We compared
the two paradigms for crystals with high bulk moduli under the same verification
budget (Fig.~\ref{fig:novelty}d,e; Supplementary
Note~\ref{si-note:route}). The bulk-modulus completion model ranked the
candidates of both paradigms. MatterSim \cite{yang2024mattersim} (Supplementary Fig.~\ref{si-fig:mlip-calibration}), the pre-registered
verifier, evaluated the top 100 of each, and DFT evaluated accepted subsets of the same
lists, both requiring a bulk modulus of at least 250~GPa and a Born-stable elastic
tensor (Section~\ref{sec:methods}; Supplementary Note~\ref{si-note:prereg}). Inverse design
(paradigm A, MatterGen conditioned on 400~GPa) returned 100 qualifying candidates under MatterSim and
\dftAhits{} of the \dftAN{} crystals evaluated by DFT, whereas forward screening
(paradigm B, the 13.52 million in-domain database structures without a recorded
bulk modulus) returned 55 under MatterSim and \dftBhits{} of \dftBN{} under DFT
(Supplementary Table~\ref{si-tab:dft-settings}).
In the original ranked lists, the largest DFT bulk modulus among the qualifying candidates is \dftAbest~GPa for inverse design
and \dftBbest~GPa for forward screening. The DFT-confirmed candidates of inverse design
fall in \dftAchem{} chemical systems, \dftAOsPhrase{} of them contain Os, and
\dftArecorded{} have a composition already recorded in the database. Within the
forward-screening list, 44 of the 55 candidates qualifying under MatterSim are carbon polytypes, and 43 of the 45
that fail the criteria belong to four families whose training labels conflict between sources or
come from a single source.

The subset of forward screening (paradigm B) drawn
from experimental structure databases (B-exp) returned 63 qualifying candidates under MatterSim,
7 of them with confirmed synthesis evidence, and \dftBexphits{} among
\dftBexpN{} under DFT. Restricted further
to structures with confirmed synthesis evidence, the top 100 of B-exp
gave 50 qualifying candidates under MatterSim and \dftBexpShits{} among \dftBexpSN{} under DFT.
Structural matching beyond the deduplication fingerprint connects at least 31 of
the top 100 forward-screening structures to entries with recorded bulk moduli (20 to
36 of their \dftBhits{} DFT-confirmed candidates, depending on the matching test),
compared with at most one in random samples of 100 forward-screening structures
(Supplementary Note~\ref{si-note:route}; Supplementary Fig.~\ref{si-fig:route-detail} and
Supplementary Table~\ref{si-tab:route}).
Across the full candidate pools, the bulk-modulus model predicts 228{,}413 structures at or above
250~GPa for forward screening, 594 of them in B-exp, compared with 1{,}074 for inverse design
(Fig.~\ref{fig:novelty}d). In a random sample of these 228{,}413 forward-screening structures (S-hi), \dftShihits{} of the \dftShiN{}
evaluated by DFT meet the criteria (\dftShirate\%).
The ordering of the two paradigms is unchanged with one structure per formula
(Supplementary Note~\ref{si-note:route}).
Moreover, the DFT-confirmed candidates span \dftAchem{} chemical systems
for inverse design, compared with \dftBexpchem{} in B-exp and \dftBexpSchem{} in its
subset with confirmed synthesis evidence, and the pre-registered MatterSim
comparison shows the same pattern in the largest bulk moduli (375.4 versus
433.8~GPa) and in chemical diversity (five versus 23 chemical systems). Inverse design yields the higher fraction of candidates meeting the criteria,
whereas forward screening reaches the larger verified bulk modulus and the
experimental-source subsets cover more chemical systems. The success fraction,
attainable property values and chemical diversity thus describe different
advantages of the two paradigms (Supplementary Note~\ref{si-note:prereg} and
Supplementary Table~\ref{si-tab:route-variants}).

\FloatBarrier

%% file: discussion.tex
\section{Discussion}

We probe the limits of forward screening and inverse design through a
comprehensive crystal database and accurate prediction of missing properties.
MateriaViva brings together 55 databases to establish 14{,}590{,}485 distinct
crystal structures and connect their previously dispersed property records.
This coverage determines both which known crystals can be screened for a target
application and which inverse-designed structures can be recognized as already
reported. Cross-database matching is essential to both questions: a structure
absent from one database can occur in another, together with calculations or
synthesis evidence that would otherwise be overlooked. Linking these records
also reveals differences between property values reported for the same crystal,
allowing their methods and sources to be compared. Because every added source extends the reference set against which subsequent
entries are matched, we paired the database with a background autonomous agent
that continuously detects newly released materials repositories and integrates
them into MateriaViva, transforming a static compilation into a self-expanding
reference that continuously updates the boundaries of both paradigms as community
data grow.

Searching this expanded collection for useful materials requires accurate
predictions of properties that previous studies did not evaluate. Our equivariant
CGTNet achieves state-of-the-art accuracy on \NbenchWins{} benchmark tasks, and its
140 property-prediction models add 874.1 million predicted values across the
database.
For the bulk modulus, completion expands coverage from 0.20\% to 93.2\% of
structures. It reveals high-performance candidates among materials already
studied computationally or synthesized experimentally whose target properties
have no recorded values. The four application examples illustrate these opportunities, with DFT
validation of selected mechanical and optical candidates supporting the
predicted performance. The potential of forward screening has therefore
been underestimated: the scientific value of a known crystal extends beyond
the properties investigated when it was first reported. Accurate property
completion and the accompanying open library of 140 pretrained CGTNet models make
this untapped potential directly accessible to systematic screening and rapid
candidate evaluation
(Supplementary Notes~\ref{si-note:discussion} and~\ref{si-note:mlip}).

For inverse design, the same structural coverage exposes the overestimated
novelty of generated crystals. Across the evaluated inverse-designed structures,
55--100\% are not new under our strictest novelty criterion, matching known
crystals in composition or in both stoichiometry pattern and space group.
Reanalyses of GNoME, A-Lab and MatterGen likewise identified materials that were
already known or closely related to reported structures
\cite{cheetham2024artificial,leeman2024challenges,juelsholt2026continued}.
A comprehensive database allows such relationships to be checked before
candidates are selected for costly verification or synthesis. Incorporating
these checks into inverse design can prevent known materials from being
presented as new discoveries and concentrate the search on structures beyond
the known crystal databases. Combined with accurate property prediction, this
provides a route to designing new structures with high performance, with novelty
and target properties considered together from the outset.

Our equal-budget comparison further shows that structural novelty and attainable
performance are separate questions. The same bulk-modulus model ranks both
candidate lists, and qualifying crystals have a bulk modulus of at least
250~GPa and a Born-stable elastic tensor in DFT. Among the evaluated candidates, inverse design
yields a higher fraction meeting these criteria than forward screening
(100\% versus 59\%), whereas screening reaches the higher verified bulk modulus
(\dftBbest{} versus \dftAbest{}~GPa). The experimental-source screening subsets,
including the subset restricted to synthesis evidence, also span more chemical
systems than generation, whose qualifying crystals all contain osmium.
Evaluating a design strategy thus requires weighing the fraction of candidates
reaching a target, the best properties attained and the chemical diversity
available for further selection. By combining cross-database deduplication with
property completion, MateriaViva turns dispersed databases into a common basis
for discovery, revealing unrecognized performance among known crystals and
directing inverse design towards new structures with useful properties. The
scope of the resource and of the comparisons drawn from it is set out in
Supplementary Note~\ref{si-note:limits}.

%% file: methods.tex
\section{Methods}
\label{sec:methods}
\subsection{Database construction and crystal deduplication}
One adapter per database extracts each entry's native identifier,
its values with units and methods, and its structure, if any
(Supplementary Table~\ref{si-tab:sources}). Property names map onto a registry
of 166 properties, and a field takes an existing name only if it denotes the same
physical quantity. To deduplicate crystals, each structure is reduced to its
primitive cell with pymatgen \cite{ong2013python} and fingerprinted with the BAWL
hasher \cite{ramlaoui2025lemat}. An entry is matched first by any
authoritative cross-database identifier it carries, then by fingerprint, and
never by chemical formula. Structures with partial occupancies are excluded
from fingerprint matching. Enumerated framework libraries are indexed with a composite
key, so that distinct frameworks sharing a bonding-graph fingerprint are not
collapsed into one entry; the fingerprint alone would merge 126{,}578 distinct Deem
zeolite frameworks into 281 structures (Supplementary Note~\ref{si-note:fingerprint}).
Any entry that remains unmatched is admitted as a new crystal, and its fingerprint
and identifiers join the matching set. Original
property values are retained without recomputation, and agreement is evaluated
where sources report the same quantity under comparable methods. For the
property-coverage analysis, tiers were assigned by hand (Supplementary
Notes~\ref{si-note:fingerprint}, \ref{si-note:audit} and~\ref{si-note:agreement};
Supplementary Table~\ref{si-tab:tiers}). Property rows are keyed on the native
identifier, property and method, and reference rows on the crystal, source and native
identifier, so that re-running a source adds new entries without duplicating existing ones.
Building on this idempotent schema, a background autonomous agent continuously
monitors the literature and public data platforms for newly released or updated
crystal repositories, automatically generating ingestion adapters to extract,
fingerprint and integrate new entries so that MateriaViva expands autonomously
over time.

\subsection{Property completion and screening}
CGTNet is specified in Supplementary Note~\ref{si-note:model} and Supplementary
Table~\ref{si-tab:gnn-config}. Each JARVIS-DFT result comes from a single run of
that configuration on the standard split. The per-task
settings and the provenance of the published values are given in Supplementary
Note~\ref{si-note:methods-model}.
For database completion, a task is one property at one level of theory, and a condition-dependent
property is split further by its conditions. Each completion task trains its own
CGTNet, holding out complete chemical systems for evaluation, and training
stops once the held-out error has not improved for
150 epochs, and the reported metrics summarize errors over the held-out set. Each trained model then predicts properties for unlabelled structures within its
applicability domain (Supplementary Notes~\ref{si-note:census}
and~\ref{si-note:completion}). Training and inference were performed on 32 NVIDIA RTX 4090-48G
and 6 NVIDIA RTX PRO 6000D GPUs. Every trained checkpoint behind the benchmark
tables and the completion library is released as a model zoo at
{\urlstyle{same}\url{https://huggingface.co/datasets/zhilong777/MateriaViva-ModelZoo}},
each with its resolved configuration, target scaler and held-out metrics
(Supplementary Note~\ref{si-note:completion}). For screening, targets and thresholds were fixed
before any candidate was counted (Supplementary Note~\ref{si-note:prereg}). For each screening criterion, the recorded value at the task's level of theory is
used where available, and the prediction otherwise. A post hoc audited screen, fixed before any audited
count, checks compound-level absence of a recorded value and coverage of the
candidate's elements and material family in the training data. The cathode screen
additionally requires a redox-active cation. Synthesis evidence identifies a
separate subset of audited candidates (Supplementary Note~\ref{si-note:screening}).

\subsection{Novelty audit and equal-budget comparison}
Generated structure sets are fingerprinted, as released or as generated, with
the same deduplication protocol and compared with the integrated crystal database.
Each structure is assigned to the strictest of the four crystal-matching criteria
under which a database entry matches it. A separate disorder-sensitivity audit
checks compositions against partially occupied experimental entries
(Supplementary Note~\ref{si-note:disorder}). For the equal-budget comparison,
the bulk-modulus completion model ranked 1{,}074 valid, fingerprint-distinct MatterGen
\cite{zeni2025generative} structures generated for a bulk modulus of 400~GPa
(inverse design, paradigm A) and the 13{,}521{,}404 database structures without a recorded bulk modulus
(forward screening, paradigm B).
 MatterSim \cite{yang2024mattersim}, the pre-registered verifier,
evaluated the top 100 of each list, and a qualifying candidate has a bulk modulus of at least
250~GPa and a positive-definite elastic tensor \cite{mouhat2014necessary}
(Supplementary Notes~\ref{si-note:route} and~\ref{si-note:mlip}). Candidates were
selected by decreasing predicted bulk modulus. For inverse design, the predicted bulk moduli of ranks 1 to 100 span only 13.7~GPa, which lies within the model's held-out
root-mean-square error of 15.2~GPa and is consistent with generation conditioned
on a single target of 400~GPa. The frozen top-100 selection was retained, and all reported verification results refer to it.

\subsection{DFT calculations}
Selected candidates were evaluated with DFT under physical settings fixed before
any result was read (Supplementary Note~\ref{si-note:mlip};
Supplementary Table~\ref{si-tab:dft-settings}). The calculations cover crystals
from the ranked lists of the equal-budget comparison and their variants, the
pre-registered and audited Prop.~1 selections, synthesized Prop.~1 and Prop.~4
candidates, structures of the pre-registered Prop.~4 pool and the two random
samples of paradigm B. The generated structures and stored database structures
serve as the starting geometries. The calculations use PBE
\cite{perdew1996generalized} with projector augmented-wave potentials
\cite{blochl1994projector,kresse1999ultrasoft} in the Vienna Ab initio Simulation
Package (VASP) \cite{kresse1996efficient}. The elastic calculations use the
700~eV cutoff and 7{,}000 $k$-points per reciprocal atom of the Materials Project
elastic data \cite{dejong2015charting} and the initial magnetic moments and
$+U$ values of the Materials Project. The elastic tensor is fitted to the
stresses of the strained cells exactly as for MatterSim, and a qualifying
candidate has a Hill bulk modulus of at least 250~GPa and a Born-stable tensor.
For Prop.~4, density-functional perturbation theory \cite{gajdos2006linear} gives
the electronic dielectric tensor at the stored geometry, and the birefringence
is the difference between the largest and smallest principal refractive indices.
Every DFT count refers to the crystals whose calculations met the convergence
criteria. Paired DFT--MatterSim statistics use the same accepted structures.

%% file: availability.tex
\bmhead{Data availability}
The database has full and open versions. The full version contains all 55
sources, 14{,}590{,}485 crystals and 131.8 million property values. All analyses
in this study use this version, which is not released as a file because several
sources restrict redistribution. The open version is available as a complete
direct download from the Hugging Face Hub at
{\urlstyle{same}\url{https://huggingface.co/datasets/AI4QC/MateriaViva-Open}},
and accordingly no query application programming
interface (API) is provided for it. It includes 10{,}903{,}588 crystals whose
licence permits redistribution and commercial use, with their structures,
provenance and per-crystal licence, together with the 113{,}459{,}302 property
values and 850{,}417 curves recorded for them, each labelled with its source and
licence; Supplementary Note~\ref{si-note:licence} sets out how a crystal
qualifies and which licence it carries.
The library of pretrained CGTNet models is publicly available on the Hugging Face
Hub as a model zoo at
{\urlstyle{same}\url{https://huggingface.co/datasets/zhilong777/MateriaViva-ModelZoo}},
each with its resolved configuration, target scaler and held-out metrics. The continuously updated database, including new
repositories ingested by the autonomous expansion agent, can be searched and
browsed at {\urlstyle{same}\url{https://materia-viva.org}}.

\bmhead{Code availability}
The acquisition adapters, structural-fingerprinting and matching pipeline,
loader, completeness audit, and data-extraction, training, analysis and
figure-generation scripts are available at
{\urlstyle{same}\url{https://github.com/AI4QC/materia-viva}}.

%% file: backmatter.tex
\bmhead{Acknowledgements}
We thank the maintainers of the 55 source databases for making their data available.

\bmhead{Funding statement}
This work was supported by the Start-up Fund of the Hong Kong University of Science and Technology.

\bmhead{Author contributions}
Z.S. and L.C. conceived the study. Z.S. built the data pipeline, developed and
trained CGTNet, performed the analyses, prepared the figures and wrote the
manuscript. L.C. supervised the study and contributed code to monitor
source-database updates, ingest properties and support the database web
interface. Z.P. redesigned the materials explorer, the interface for searching
and viewing the database.

\bmhead{Competing interests}
The authors declare no competing interests.

\bmhead{Additional information}
Supplementary Information accompanies this manuscript. Correspondence should be addressed to Lixue Cheng (\href{mailto:lixuecheng@ust.hk}{lixuecheng@ust.hk}).